\documentclass[twocolumn]{aastex61}
\usepackage{graphicx}
\usepackage{amsbsy}
\usepackage{amsmath} 
\usepackage{amssymb}
\usepackage[utf8]{inputenc}
\usepackage[toc,page]{appendix}
\usepackage[T1]{fontenc}
\usepackage{hyperref}
\usepackage{color}

\submitjournal{AJ}

\shorttitle{}
\shortauthors{Cambioni \& Malhotra}

\begin{document}
\bibliographystyle{plainnat}
\title{The mid-plane of the Main Asteroid Belt}

\correspondingauthor{Saverio Cambioni}
\email{cambioni@lpl.arizona.edu}

\author{Saverio Cambioni}
\affil{Lunar and Planetary Laboratory, 1629 E University Blvd, Tucson, AZ, 85721-0092, USA}

\author{Renu Malhotra}
\affil{Lunar and Planetary Laboratory, 1629 E University Blvd, Tucson, AZ, 85721-0092, USA}

\begin{abstract}
We measure the mid-plane of the main asteroid belt by using the observational data of a nearly complete and unbiased sample of asteroids, and find that it has inclination $\bar{I}=0.93\pm0.04$ degrees and longitude of ascending node $\bar{\Omega}=87.6\pm2.6$ degrees (in J2000 ecliptic-equinox coordinate system). This plane differs significantly from previously published measurements, and it is also distinctly different than the solar system's invariable plane as well as Jupiter's orbit plane. The mid-plane of the asteroid belt is theoretically expected to be a slightly warped sheet whose local normal is controlled by the gravity of the major planets. Specifically, its inclination and longitude of ascending node varies with semi-major axis and time (on secular timescales), and is defined by the forced solution of secular perturbation theory; the $\nu_{16}$ nodal secular resonance is predicted to cause a significant warp of the mid-plane in the inner asteroid belt. We test the secular theory by measuring the current location of the asteroids' mid-plane in finer semi-major axis bins. We find that the measured mid-plane in the middle and outer asteroid belt is consistent, within 3--$\sigma$ confidence level, with the prediction of secular perturbation theory, but a notable discrepancy is present in the inner asteroid belt near $\sim2$ AU.
\end{abstract}

\section{Introduction} \label{s:intro}

The orbital distribution in the main asteroid belt is widely recognized to hold clues to the formation and dynamical evolution of the solar system’s planets (e.g., \cite{Obrien:2011, Morbidelli:2015}), and many important non-trivial features have been identified in the dynamical structure of the asteroid belt (e.g., \cite{Nesvorny:2002, Michtchenko:2016, Malhotra:2017}).  However, one important dynamical property that has received relatively little attention is the mid-plane of the asteroid belt.  This mid-plane is controlled by the gravity and the angular momenta of the planets, including possible undiscovered distant planets and other transient perturbations (such as recent stellar encounters) that may affect the barycenter and angular momentum of the solar system.
The earliest attempt to determine the mid-plane of the observed main belt asteroids (MBAs) appears to have been by \cite{Plummer:1916}. He examined the available orbital data of 753 minor planets in order to deduce the law of distribution of their orbital parameters. 
The mean plane was found to have inclination of $0.89$ degrees and longitude of ascending node of $106.7$ degrees, relative to the ecliptic, and the orbital poles of the minor planets were measured to have a root mean square deviation of $11.2$ degrees about the mean pole.
Importantly, \cite{Plummer:1916} also mentioned the observational circumstances as one of the factors to be considered in the interpretation of the results, which were later examined by \cite{Kresak:1989}. 

In the following years, a few studies carried out analyses of the distribution of inclination and longitude of ascending node without reporting the mean plane  \citep{Michkovitch:1948, Zheverzheev:1979}. In the last two decades of the 20th century, the spread of Charge-Coupled Devices (CCDs) for astronomical observations as well as increased attention to the potential impact hazard represented by near-Earth asteroids produced a large increase in the rate of minor planet discoveries \citep{Stokes:2002}. \cite{Shor:1992} reported that the mean plane of the main belt population had inclination $1.00$ degrees and its longitude of ascending node was $96.4$ degrees, relative to the ecliptic; no uncertainty was reported for these measurements which were based on a sample of 1193 MBAs of mean opposition apparent magnitude below 14.5.\\

Since the time of the previous studies, the number of known MBAs has dramatically increased, accounting today for a population of almost 500,000 members. The large increase of the MBA observational sample motivates us to revisit the problem of the determination of the mean plane of the asteroid belt. The large sample size allows us to constrain the overall mean plane of the population at much higher accuracy than in previous studies. We also measure the mean plane as a function of semi-major axis by binning the population and computing the local mean plane, for comparison with theoretical predictions. Throughout this paper, we refer all orbital elements to the J2000 heliocentric ecliptic-equinox coordinate system. The rest of this paper is organized as follows. The theoretically predicted mid-plane is described in Section \ref{s:theory}. The methods for measuring the mean plane of a population of objects and, importantly, its measurement uncertainty are presented in Section \ref{s:method1} and Section \ref{s:method2} respectively.  The observational dataset is described in Section \ref{data}. The results are presented in Section \ref{results}, and we discuss their implications in Section \ref{s:discussion}.
\section{Expected mean plane}
\label{s:theory}

The masses of individual minor planets in the Main Asteroid Belt are negligible compared to those of the major planets, therefore we can consider an asteroid as a test particle. 
The Laplace-Lagrange linear secular theory \citep{Murray:1999} provides an analytical solution for the orbit-averaged secular motion of a test particle whose orbit around the Sun is perturbed by the eight planets. Each of the eight planets perturb the orbit of each test particle and also each other according to the secular terms in the disturbing function. These terms describe the long term perturbations from the planets, which strongly influence the orbital distribution of the minor planets by enforcing their mean plane. The osculating orbit of a test particle has inclination vector that can be decomposed as the vector sum of a “proper inclination” vector (also known as “free inclination”) and a “forced inclination” vector, 
\[\sin{I}(\cos{\Omega},\sin{\Omega}) =
\sin{I_{\mathrm{p}}}(\cos{\Omega_{\mathrm{p}}},\sin{\Omega_{\mathrm{p}}}) \]
\[+\sin{I_0}(\cos{\Omega_0},\sin{\Omega_0})\] 
The proper inclination vector depends on the initial conditions of the test particle, while the forced inclination vector depends upon the gravitational perturbations of the major planets. In the linear secular theory, the forced inclination vector of test particles perturbed by the planets is given by,
\begin{equation}
(\sin{I_0}\cos{\Omega_0},\sin{I_0}\sin{\Omega_0})=\sum_{i=1}^8 \frac{\mu_i}{f_i-f_0}(\cos{\gamma_i},\sin{\gamma_i}),
\label{e:forcedincvec}\end{equation}
where $\mu_i$ are weighting factors for each secular mode of the eight planets, $f_0$ is the nodal precession rate of the test particle induced by the orbit-averaged quadrupolar potential of the planets, $f_i$ and $\gamma_i$ are the frequencies and phases of the nodal secular modes of the Solar System's eight major planets. The parameters $f_i$ and $\gamma_i$ depend only on the planetary parameters, while $f_0$ and $\mu_i$ depend additionally on the test particle's semi-major axis. As a result, the forced inclination $I_0$ and longitude of ascending node $\Omega_0$ of a test particle are a function of the minor planet's semi-major axis. 

It is worth noting that the secular solution from the linear perturbation theory is formally valid in the low-inclination regime and for asteroids whose semi-major axes are well separated from the planets'. An additional condition is that the asteroids must be away from strong mean-motion resonances. The free inclination vector does not have a preferred directionality, whereas the forced inclination vector is controlled by the gravitational perturbations of the planets. Consequently, for a population of minor planets with some dispersion in orbital planes and a small dispersion in semi-major axis, the mean plane will coincide with the secularly forced plane given by the forced inclination (Eq.~\ref{e:forcedincvec}) at the epoch of the osculating elements of the planets. Indeed, as the planets' orbits evolve under their mutual perturbations over secular timescales, the forced inclination vector changes on timescales $>10^4$ years for the MBAs. At epoch February 16, 2017, the values of $I_0$ and $\Omega_0$ for the semi-major axis range 2.00--3.30 AU are reported in Figure \ref{a_warps} as the blue and red curves.  
The blue curves plot the forced inclination vector from the linear secular theory by following the analytical procedure described in \cite{Murray:1999} and using the masses, semi-major axes, inclinations and longitudes of ascending nodes for the eight major planets, Mercury--Neptune, from JPL Horizons (\url{https://ssd.jpl.nasa.gov/horizons.cgi}). (The secular mode frequencies and amplitudes of the planets that we obtained are very similar to the values tabulated in \cite{Murray:1999}; the maximum fractional difference for the eight secular mode frequencies is less than 2\%, the maximum fractional difference for the amplitude of the fastest mode, $f_6$, is 1.1\%.) The red curves in Figure \ref{a_warps} plot the forced inclination vector based on the numerically computed secular theory for the planets \citep{Morbidelli:2002}. These numerically computed mode frequencies and amplitudes differ more significantly from those of the linear theory.

We observe in Figure \ref{a_warps} that the predicted mean plane is not flat across the whole asteroid belt, but changes with heliocentric distance, and it has a prominent warp in the inner belt; this warp is due to the $\nu_{16}$ nodal secular resonance \citep{Knezevic:1991}. As the semi-major axis increases, both the forced inclination and longitude of ascending node achieves a plateau, approaching the respective value for the orbit of Jupiter. This is a general result of the linear secular perturbation theory, for which the forced elements at a planet's location coincide with the corresponding osculating elements of the planet \citep{Murray:1999}.
\section{Measuring the mean plane} \label{s:method1}
In J2000 ecliptic/equinox, the components of the unit vector directed toward the orbital pole of a heliocentric Keplerian orbit are defined in terms of the ecliptic inclination, $I$,  and the longitude of ascending node, $\Omega$: 
\begin{equation}
\boldsymbol{\hat{n}}=(\sin{I}\sin{\Omega},-\sin{I}\cos{\Omega},\cos{I}).
\end{equation}
For a sample of dispersed $N$ orbit planes, the unit vectors $\boldsymbol{\hat{n}_i}$ are constrained to a sphere of unit radius, because every vector has unit magnitude. The mean pole of the population is computed as: 
\begin{equation}
\boldsymbol{{n}_{av}}=(\frac{\sum_i^N{\hat{n}_{1,i}}}{N},~\frac{\sum_i^N{\hat{n}_{2,i}}}{N},~\frac{\sum_i^N{\hat{n}_{3,i}}}{N}).
\end{equation}
The resulting average vector must be normalized to get the unit vector of the mean pole of the sample:
\begin{equation}
\boldsymbol{\bar{n}}=\frac{\boldsymbol{{n}_{av}}}{||\boldsymbol{n_{av}}||}.
\end{equation}
This unit mean pole vector describes the mean plane whose inclination and longitude of ascending node are retrieved as follows: 

\begin{equation}
\bar{I}=\arccos({\bar{n}_3}),~~~\bar{\Omega}=\arctan\bigg(-\frac{\bar{n}_1}{\bar{n}_2}~\bigg)
\end{equation}

\section{Measurement uncertainty of the mean plane} \label{s:method2}

The measurement uncertainty of the mean plane is estimated by deriving the distribution of the mean poles with bootstrapping simulations \citep{Efron:1979}. The bootstrapping method allows assigning measures of accuracy to sample estimates through a resampling procedure as follows.
From the original sample of orbital poles $[\boldsymbol{\hat{n}}_1,~\boldsymbol{\hat{n}}_2,...,\boldsymbol{\hat{n}}_N]$, a sample $[\boldsymbol{\hat{n}}_1^*,~\boldsymbol{\hat{n}}_2^*,...,\boldsymbol{\hat{n}}_N^*]$ is drawn at random, with replacement; the mean pole of the bootstrapped population is computed by averaging the orbital poles. This resampling procedure is repeated many times (typically 1000-10000) in order to get a good statistics of the bootstrapped estimator $\boldsymbol{\bar{n}}^*$. 
The bootstrapped population of mean poles is then projected onto the ecliptic plane in order to estimate the measurement uncertainty of the mean plane's ecliptic inclination $\bar{I}$ and longitude of ascending node $\bar{\Omega}$. On the ecliptic, the projected mean poles have components:
\begin{equation}
({\bar{n}_1}^*,{\bar{n}_2}^*)=(\sin{I}\sin{\Omega},-\sin{I}\cos{\Omega}) \equiv (P,-Q),
\end{equation}
where we have introduced the usual ``inclination vector" components $P$ and $Q$ \citep{Murray:1999}, which are related to the unit vector normal to the orbit plane.
The uncertainties in the ecliptic inclination and longitude of ascending node are computed graphically by considering the distribution of mean poles' projections in the $(Q,P)$ plane, computing their 1-$\sigma$ contour, and then locating the extrema of $I$ and $\Omega$ on the 1--$\sigma$ contour.

\begin{figure*}
\plottwo{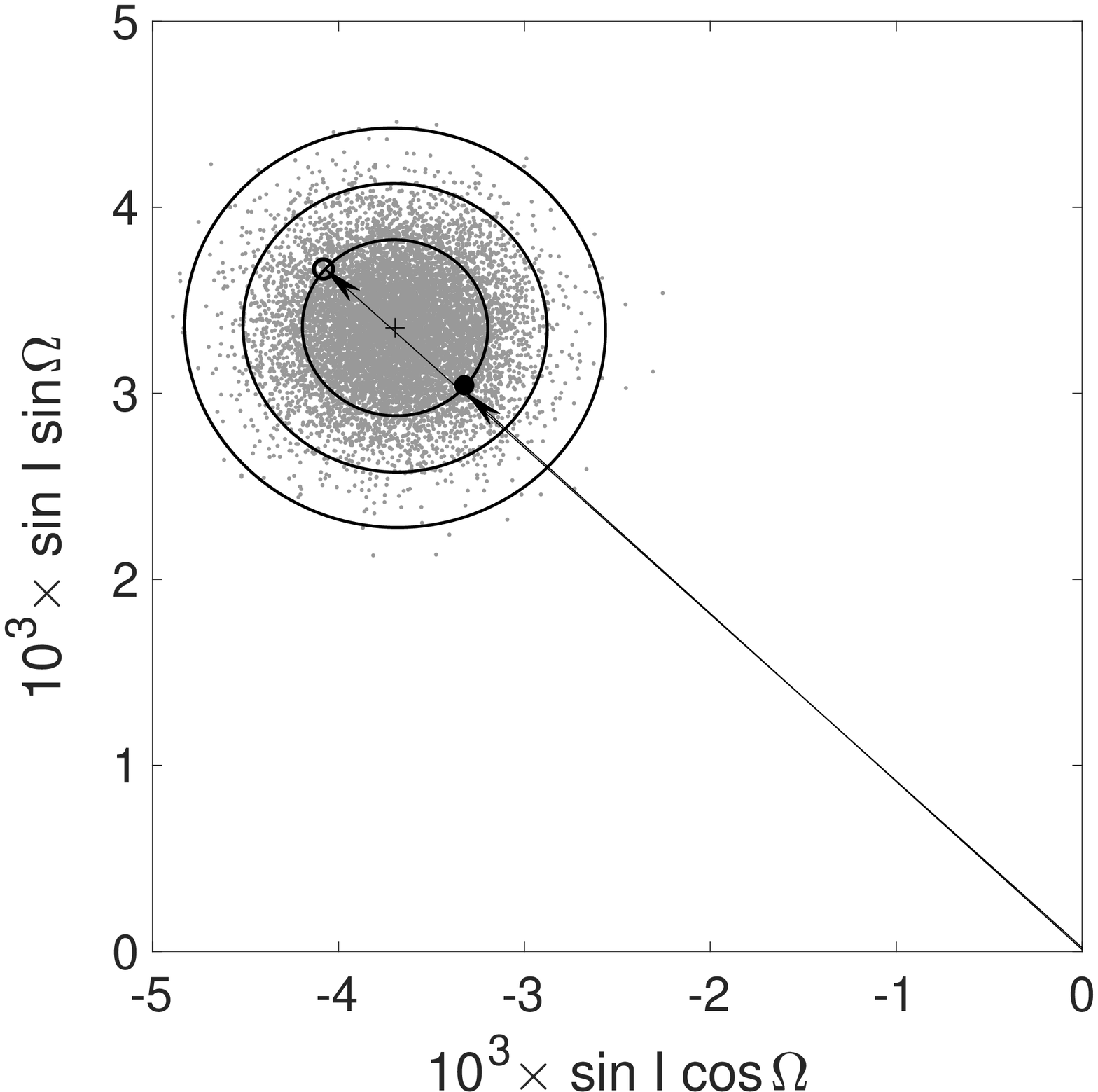}{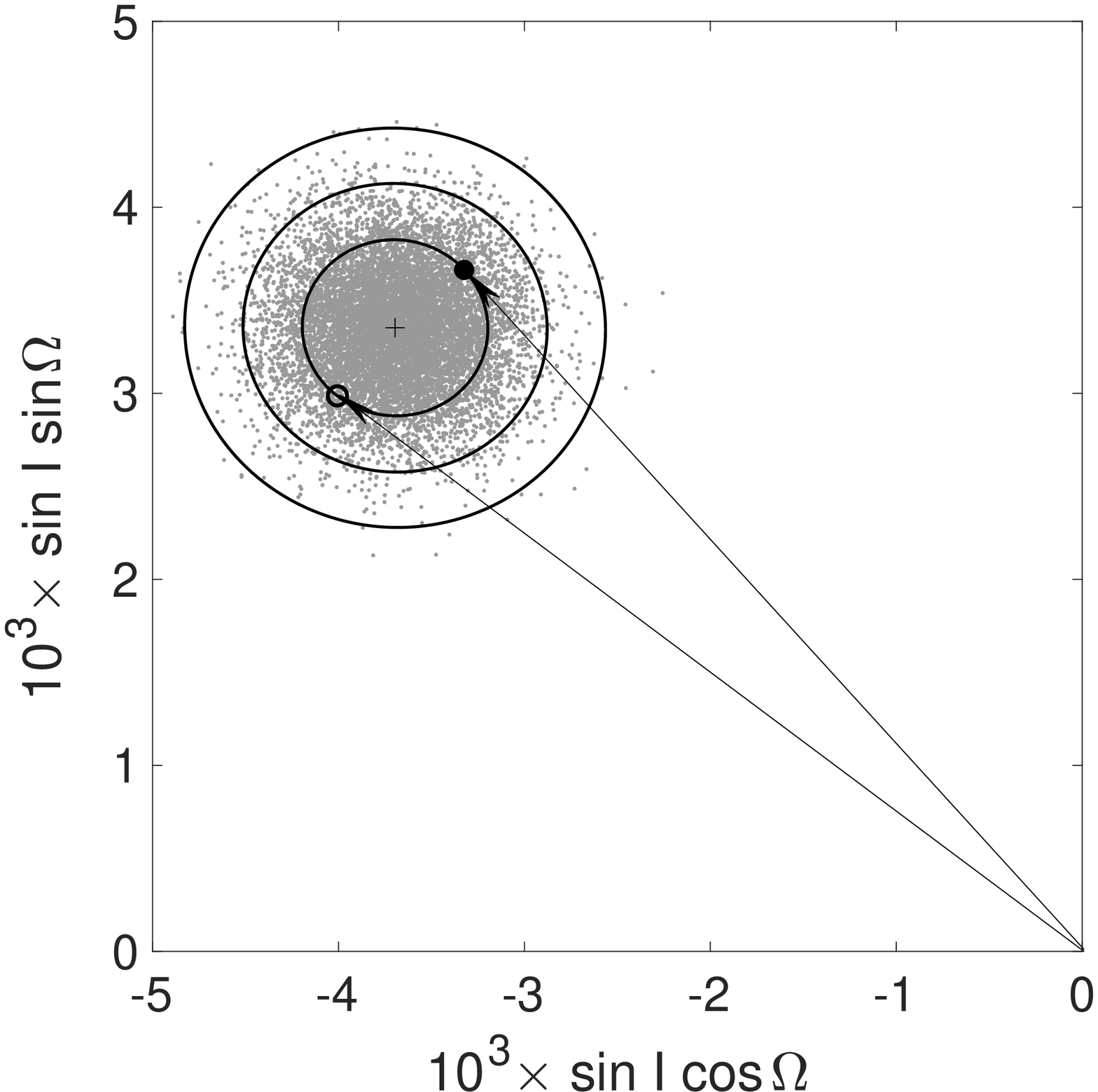}
\centering
\caption{Illustration of our method for estimating the measurement uncertainty of the mean pole. The bootstrapped mean poles are projected onto the ecliptic $(Q,P)$ plane. The black cross indicates the prescribed mean pole, while the ecliptic pole is  the bottom-right corner of the panels. In the left panel, the arrows indicate the points on the error contour for which the difference in inclinations with respect to the measured mean pole is maximum. In the right panel, the arrows indicate the points on the error contour for which the difference in longitude of ascending node with respect to the mean pole is maximum. The black contours are ellipses enclosing the 68.3\%, 95.4\% and 99.7\% fractions of the sample (1--$\sigma$, 2--$\sigma$, 3--$\sigma$ contours respectively).}
\label{unc}
\end{figure*}

This method is illustrated in Figure \ref{unc} for a simulated sample of bootstrapped mean poles. The synthetic population of Figure \ref{unc} has been generated for an example to illustrate to the reader the geometrical procedure for the estimation of the uncertainty. We generated a synthetic sample of 10000 orbit planes
from a Gaussian distribution of values of $P$ and $Q$, with prescribed mean values
$\bar P = 0.003$ and  $\bar Q =-0.003$ and standard deviation 0.01 and measured the sample's mean pole. Then we computed the distribution of mean poles by means of bootstrapping with replacement; these are plotted as the gray dots in Figure~\ref{unc}. The black cross indicates the prescribed mean pole, while the ecliptic pole is  the bottom-right corner of the panels. The concentric black contours are the ellipses enclosing the 68.3\%, 95.4\% and 99.7\% fractions of the sample (1--$\sigma$, 2--$\sigma$, 3--$\sigma$ contours respectively). 
In the left panel of Figure~\ref{unc}, the arrows indicate the points on the 1--$\sigma$ contour for which the difference in ecliptic inclinations with respect to the mean pole is maximum. $I_{-}$ is the inclination of the filled dot (closest to the ecliptic pole), while $I_{+}$ is the inclination of the open dot (farthest from the ecliptic pole). Consequently, the 1--$\sigma$ uncertainties of the inclination are given by
\begin{equation}
\sigma_{I,+}=I_{+}-\bar{I},~~~~~~\sigma_{I,-}=\bar{I}-I_{-}
\end{equation}
In the right panel of Figure~\ref{unc}, the arrows indicate the points on the 1--$\sigma$  contour for which the difference in longitude of ascending node with respect to the mean pole is maximum; the angles are measured positive counterclockwise from the equinox axis. Thus, $\Omega_{-}$ is the value of the ascending node which is angularly closest to the equinox axis, indicated by the filled dot, while $\Omega_{+}$ is the value of the ascending node which is angularly farthest to the equinox axis, and is indicated by the open dot. Consequently, the 1--$\sigma$ uncertainties of the node are given by
\begin{equation}
\sigma_{\Omega,+}=\Omega_{+}-\bar{\Omega},~~~~~~\sigma_{\Omega,-}=\bar{\Omega}-\Omega_{-}.
\end{equation}
Finally, the measured ecliptic inclination and longitude of ascending node of the mean plane of the population is reported as
\begin{equation}
I_m=\bar{I}^{+\sigma_{I,+}}_{-\sigma_{I,-}}~~~~~~\Omega_m=\bar{\Omega}^{+\sigma_{\Omega,+}}_{-\sigma_{\Omega,-}}.
\end{equation}
%
\section{MBA observational data}
\label{data}

We obtained the asteroids' data from the catalog available at the AstDys-2 website\footnote{\url{http://hamilton.dm.unipi.it/astdys/}}, which are updated daily.  Our analysis is based on the catalog of February 16, 2017.  The orbital elements at epoch JD 2457800.0 are given in the J2000 ecliptic/equinox coordinate system centered at the Sun. Our criteria for selecting the sample of MBAs are based on the osculating semi-major $a$ and eccentricity $e$ of the objects:
\begin{itemize}
\item $a\geq 1.60~AU$ and $a\leq 3.30~AU$
\item perihelion distance: $a(1-e)\geq 1.30~AU$
\end{itemize}
Of the sample of MBAs so-defined, we further selected down to minimize biases that would affect the measurement of the mean plane.  These biases arise due to the presence of collisional families and due to observational incompleteness.  If we are successful at this, then our measurement of the Main Belt's mean plane should be affected by random error only, which can be estimated as described in Section \ref{s:method2}.

A first systematic error arises from the presence of collisional families within the catalog. A collisional family is a group of asteroids that are thought to have a common origin in an impact (collision) which produced many fragments. Such groups have similar spectral colors and similar albedos, and share similar proper orbital elements. As a result, the members of a collisional family have their orbital planes correlated with one each other, potentially inducing a systematic error in the  MBAs' mean plane measurement. We elected to remove from our data sample all the family members save for the family's "mother" (which is identified as the largest or brightest object of the group). We used the list of collisional family members available on the AstDys-2 website. This list is the product of previous works that investigated the proper elements of asteroids in order to identify their common collisional genesis \citep{Milani:2014,Knezevic:2003}.

\begin{figure}
	\includegraphics[width=1\linewidth]{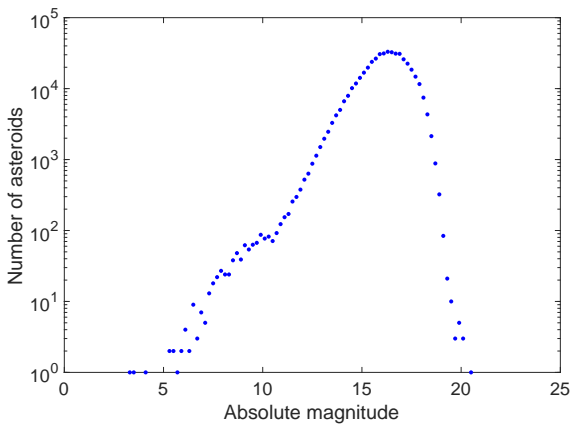}
    \caption{The distribution of the absolute magnitude of the numbered asteroids. The bin size is 0.2.}
	\label{Nvsmagn}
\end{figure}

Another bias arises from the incompleteness of the observational sample because smaller/fainter objects are harder to detect. For a nearly power-law distribution of sizes of objects, the logarithm of the number of asteroids in the Main Belt is expected to increase almost linearly with absolute magnitude. However, the observational sample suffers from the selection effect that brighter objects are more easily detectable than fainter objects. The higher detectability of bright, large objects results in an artificial peak in the population curve as a function of absolute magnitude. This is evident in Figure \ref{Nvsmagn} which plots the distribution of the absolute magnitude of the whole data sample of MBAs.  This observational incompleteness produces the peak in the curve at about magnitude 16. Since the peak is due to the limited observational capability in detecting faint/small objects, we can conclude that the observational sample above the peak magnitude is incomplete. For our purpose of measuring the mean plane of the MBAs, a conservative approach for defining a fair sample of the population is limiting the absolute magnitude of the sample to a cut-off magnitude value before the peak, say between absolute magnitude $H=14$ and $H=16$. 

In order to finalize our choice for the cut-off magnitude, we must consider that the absolute magnitudes of the AstDys2 catalog are affected from uncertainties which depend on the photometric records of the objects. A catalog of absolute magnitude uncertainties for the numbered asteroids is available on the PSR webpage of the University of Helsinki (\url{https://wiki.helsinki.fi/display/PSR}). Unsurprisingly, we find that the absolute magnitudes of fainter asteroids have higher uncertainties than those of the brighter objects because the detections of these objects are less frequent than those of brighter objects. Furthermore, for faint objects the signal to noise ratio is often too low for good determination of magnitudes.  Additional systematic errors are possibly due to the procedure of the Minor Planet Center to incorporate photometry from different sites and filters over a long period of time \citep{Veres:2015}. The typical uncertainty in the measured absolute magnitudes of asteroids is about 0.1.
For asteroids brighter than $H=16$, a significant fraction ($\sim$~32\% 
of the population devoid of collisional family members) has an uncertainty in brightness greater than 0.1; for asteroids brighter than $H=15.5$, a smaller fraction ($\sim$~ 20\%
of the population devoid of collisional family members) has an uncertainty in brightness greater than 0.1, and only $\sim$~0.45\% of the population has an uncertainty in brightness greater than 0.5.
Taking this into account, we choose a cut-off magnitude for our sample equal to $H=15.5$, getting a "margin of safety" of about $\delta H = 0.5$ from the observational incompleteness boundary. This
choice does not affect our results. We confirmed, by bootstrapping simulations, that 
the mean pole of the sample with $H<16$ and the mean pole of the sample with our chosen value of $H<15.5$ 
deviate from each other by less than 3--$\sigma$; the mean poles of the samples with cut off magnitudes $H=14.5$, $H=15.0$ are also within 3--$\sigma$, but the sample with a fainter cut-off magnitude, $H=16.5$, deviates by more than 3--$\sigma$.

\begin{figure}[htpb]
\includegraphics[width=\linewidth]{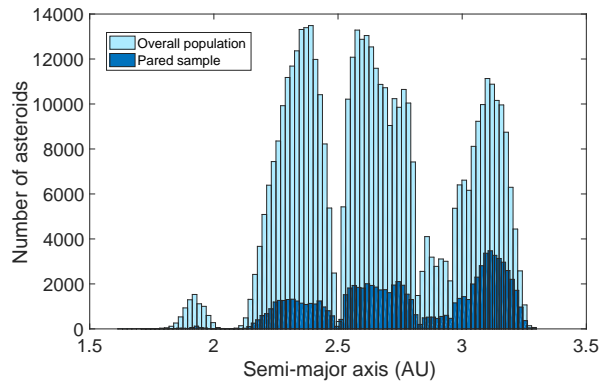}
\centering
\caption{A histogram of the number of asteroids in the overall and pared population {as a function of semi-major axis; the bin-size is 0.02 AU}.
}
\label{histo_red}
\end{figure}

After applying the above criteria, the subset of MBAs for the purpose of measuring the mean plane of the main belt consists of 89,216 objects. The semi-major axis distribution of this sample of MBAs is compared to that of the overall population of MBAs in Figure~\ref{histo_red}. The bins in semi-major axis are 0.02 AU wide. Interestingly, we note that the pared sample has a smaller fraction of members in the inner belt compared to the overall (non-pared) population; this peculiar distribution is the result of the combined effects of dynamical sculpting of the MBAs over gigayear timescales by the planets and the effects of resonance sweeping during the early migration of Jupiter and Saturn~\citep{Minton:2009}.

The minimum ecliptic inclination in the pared sample is 0.02 degrees and the maximum is 52.01 degrees. The mode of the ecliptic inclinations is near 6 degrees, and more than 80\% of the sample has ecliptic inclination less than 15 degrees. It is possible that there is observational incompleteness of the higher inclination asteroids, but the incompleteness level is not likely to be very significant for the brighter asteroid sample (absolute magnitude $H<15.5$).

We observe that there are no asteroids in our sample with semi-major axis below 1.70 AU. Between 1.70 AU and 2.10 AU, the total size of the sample (431 asteroids) is significantly smaller than in each of the other bins. The asteroids at this location are also highly dispersed in inclination, ranging from 4.48 degrees to 47.93 degrees.  

\section{Results}
\label{results}

\subsection{The overall mean plane of the main asteroid belt}

\begin{figure}
\includegraphics[width=1\linewidth]{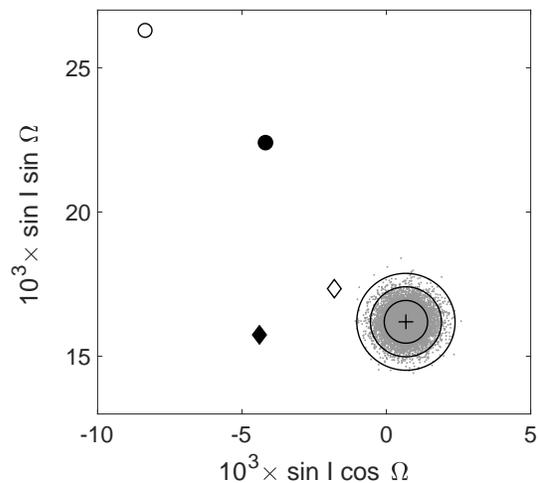}
\caption{The mean pole (black cross) of the pared sample of MBAs is indicated by the black cross; the gray dots are the bootstrapped population of mean poles projected onto the $(Q,P)$ plane. The concentric black contours are the ellipses containing the 68.3\%, 95.4\% and 99.7\% of the population, respectively. For comparison, we report measurements of the mean pole by \cite{Plummer:1916} (filled diamond) and \cite{Shor:1992} (open diamond). The open dot and the filled dot are the pole of the invariable plane and Jupiter's orbital pole respectively.}
\label{mbaunc}
\end{figure}

We measured the mean plane of our pared sample of MBAs and computed its measurement uncertainty, as described in section~\ref{s:method1} and section~\ref{s:method2}. The projection of the mean pole onto the ecliptic $(Q,P)$ plane is indicated by the black cross in Figure~\ref{mbaunc}.
 The bootstrapped population of 10,000 mean poles is also plotted in this figure, together with three concentric contours which are the ellipses containing the 68.3\%, 95.4\% and 99.7\% of the population (1--$\sigma$, 2--$\sigma$, 3--$\sigma$ contours respectively). 

We find that, in the J2000 ecliptic/mean equinox reference frame, the inclination and longitude of ascending node of the mean pole and their measurement uncertainties are as follows:
$$ \bar{I}=0.93\pm 0.04~{\hbox{degrees,}}$$
$$\bar{\Omega}=87.6\pm 2.6~{\hbox{degrees.}}$$
The previously published measurements of the mean pole by \cite{Plummer:1916} and \cite{Shor:1992}, propagated to the J2000 ecliptic-equinox coordinate system, are also shown in Figure~\ref{mbaunc} as the filled diamond and the open diamond; they can be ruled out at more than 3--$\sigma$ confidence level. These previous estimates suffered from observational incompleteness of the analyzed sample and the bias due to the presence of collisional family members in the population, which increases the degree of correlation in the data.  

The solar system's invariable plane, with inclination 1.58 degrees and longitude of ascending node 107.6 degrees~\citep{Souami:2012} has coordinates $(Q,P)=(-8.34,26.28)\times10^{-3}$, and Jupiter's orbital plane has $(Q,P)=(-4.17,22.39)\times10^{-3}$ (shown as the open dot and filled dot respectively, in Fig.~\ref{mbaunc}). The measured mean pole deviates from both of these at much more than 3-$\sigma$ confidence level. In other words, the measured mean plane of the main belt asteroids is distinctly different than the invariable plane of the solar system and also distinctly different than Jupiter's orbital plane.
\subsection{The warped mean plane}

\begin{figure*}
\includegraphics[width=1\textwidth]{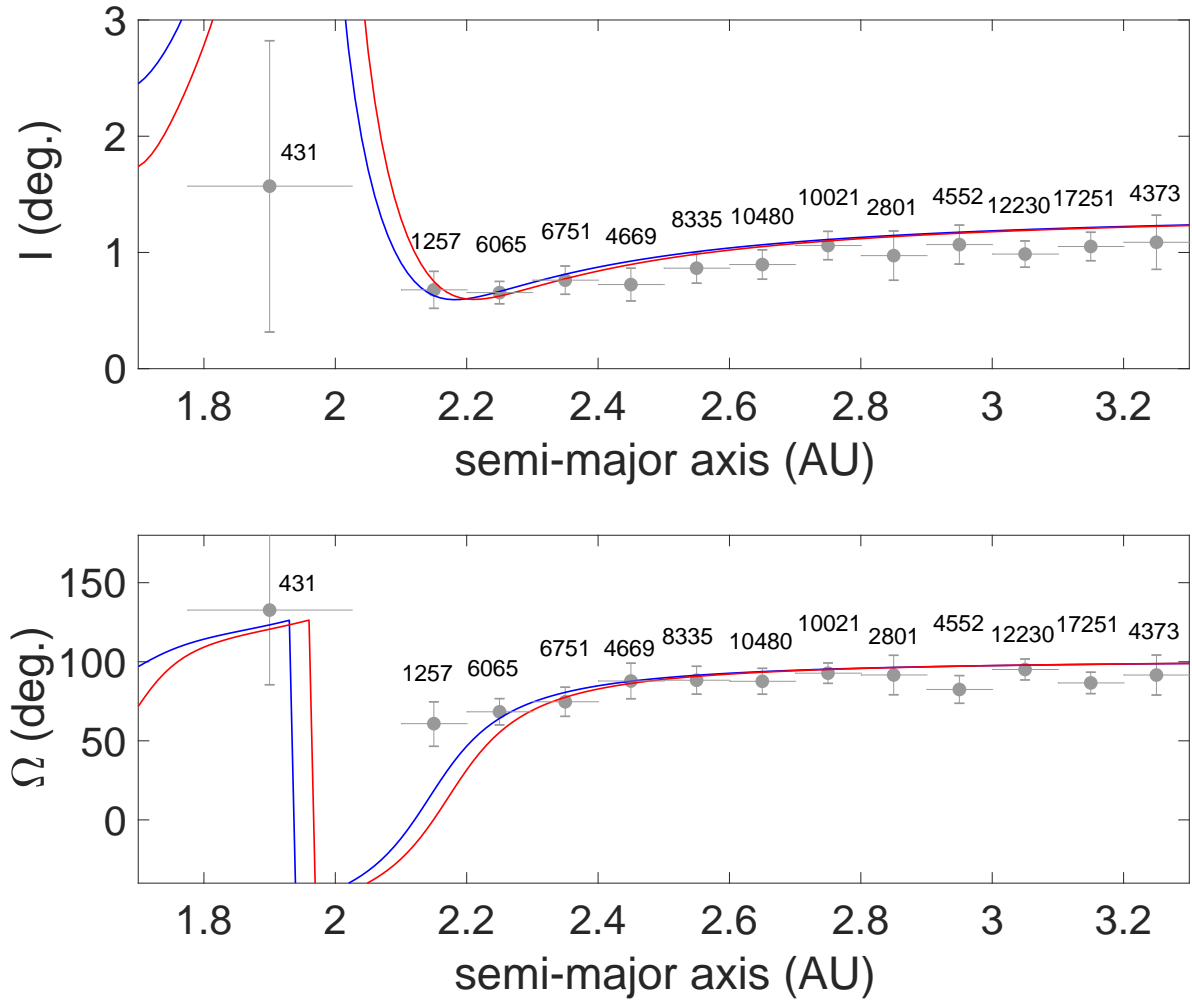}
\centering
\caption{The ecliptic inclination (top panel) and longitude of ascending node (bottom panel) of the mean plane of the MBAs as a function of semi-major axis. The gray dots plot the measured mean plane in semi-major axis bins. The vertical error bars indicate the measurement errors, and the horizontal error bars indicate the bin-size; the numbers near each point indicate the sample size in each bin. The blue curves plot the forced inclination vector computed with the linear secular perturbation theory. The red curves plot the forced inclination vector based on the numerically computed secular modes of the planets.}
\label{a_warps}
\end{figure*}

As we described in Section~\ref{s:theory}, the expected mid-plane of the main asteroid belt is not a flat sheet, but has a prominent warp near the inner edge (near semi-major axis $a\approx2.00$~AU), owing to the $\nu_{16}$ nodal secular resonance; outward of this warp, the expected mid-plane in the outer asteroid belt asymptotically approaches Jupiter's orbit plane (see Figure~\ref{a_warps}). In order to test this prediction of the secular theory, we bin the MBA sample in semi-major axis between 2.10 AU and 3.30 AU. (The sample in the semi-major axis range 1.70--2.10 AU is considered separately at the end of this section.) After some experimentation, we chose equal-sized bins of width 0.1 AU between 2.10 AU and 3.30 AU. This choice is based both upon the behavior of the secular solution and the sample size.  In Figure \ref{a_warps} we can clearly recognize that the forced inclination vector varies more rapidly in the inner belt (let say before 2.50 AU) than it does in the middle and outer belt. This observation would encourage thinner bins in the inner belt region in order to try to resolve the rapid variation, whereas the resolution in the middle and outer belt could be coarser since the forced inclination varies more smoothly. On the other hand, Figure \ref{histo_red} shows that the bins in the inner belt are much less populated than those in the middle and outer belt. Choosing thinner bins in the inner region is not useful since it leads to large standard errors in the retrieved elements of the local mean plane. 

\begin{figure*}[htpb]
\begin{center}
\includegraphics[width=.9\textwidth]{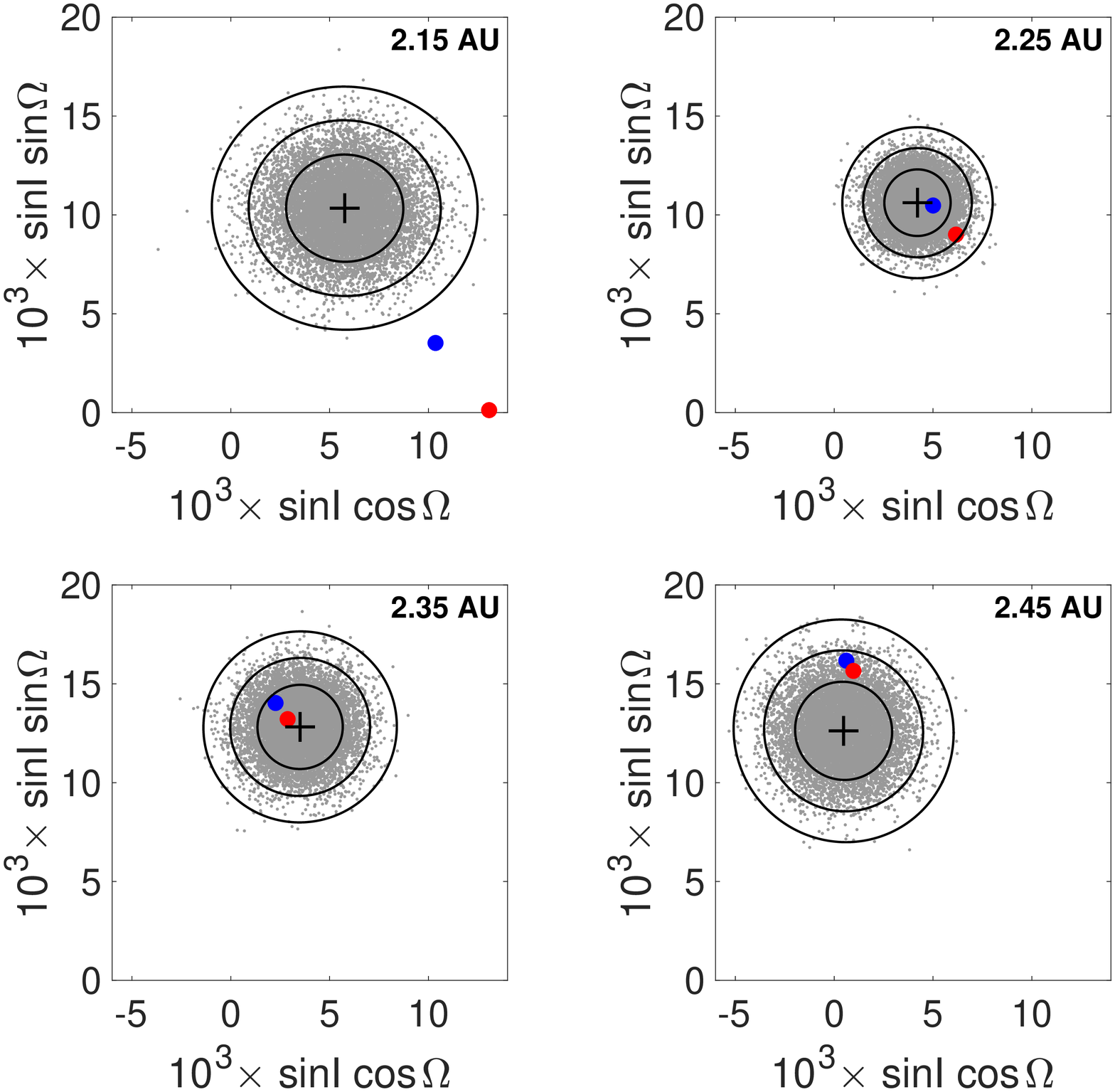}
\end{center}
\caption{The MBAs' measured mean poles (black crosses) and their predicted mean poles (blue dots and red dots for the linear theory and numerical theory, respectively) in the inner belt, in semi-major axis bins of 0.1 AU width, projected on the $(Q,P)$ plane.  The bootstrapped mean poles of each observational sample of MBAs are plotted as the gray dots, and the concentric ellipses indicate the 1--, 2-- and 3--$\sigma$ measurement uncertainty contours.  The semi-major axis bin center is indicated in each panel.}
\label{inner}
\end{figure*}
\begin{figure*}[htpb]
\begin{center}
\includegraphics[width=.9\textwidth]{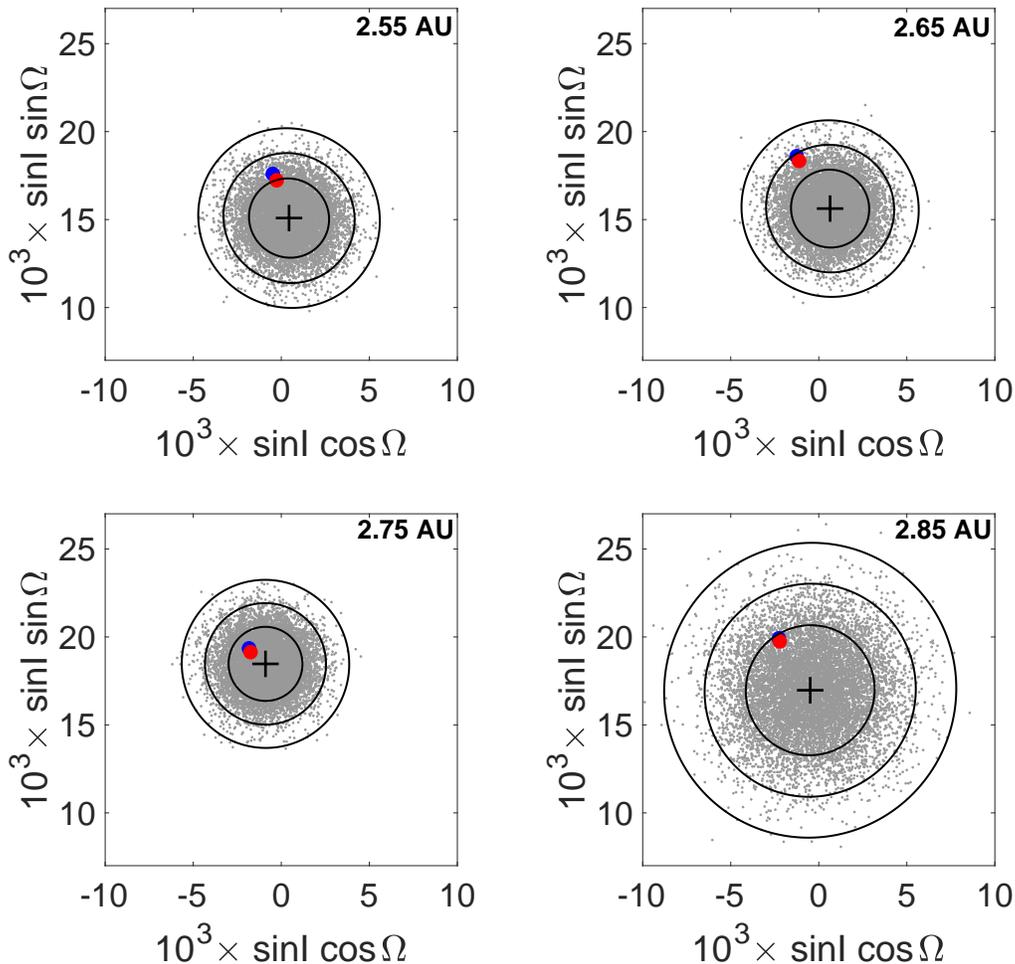}
\end{center}
\caption{Similar to Figure~\ref{inner}, but for the middle belt MBAs.}
\label{middle}
\end{figure*}
\begin{figure*}[htpb]
\begin{center}
\includegraphics[width=.9\textwidth]{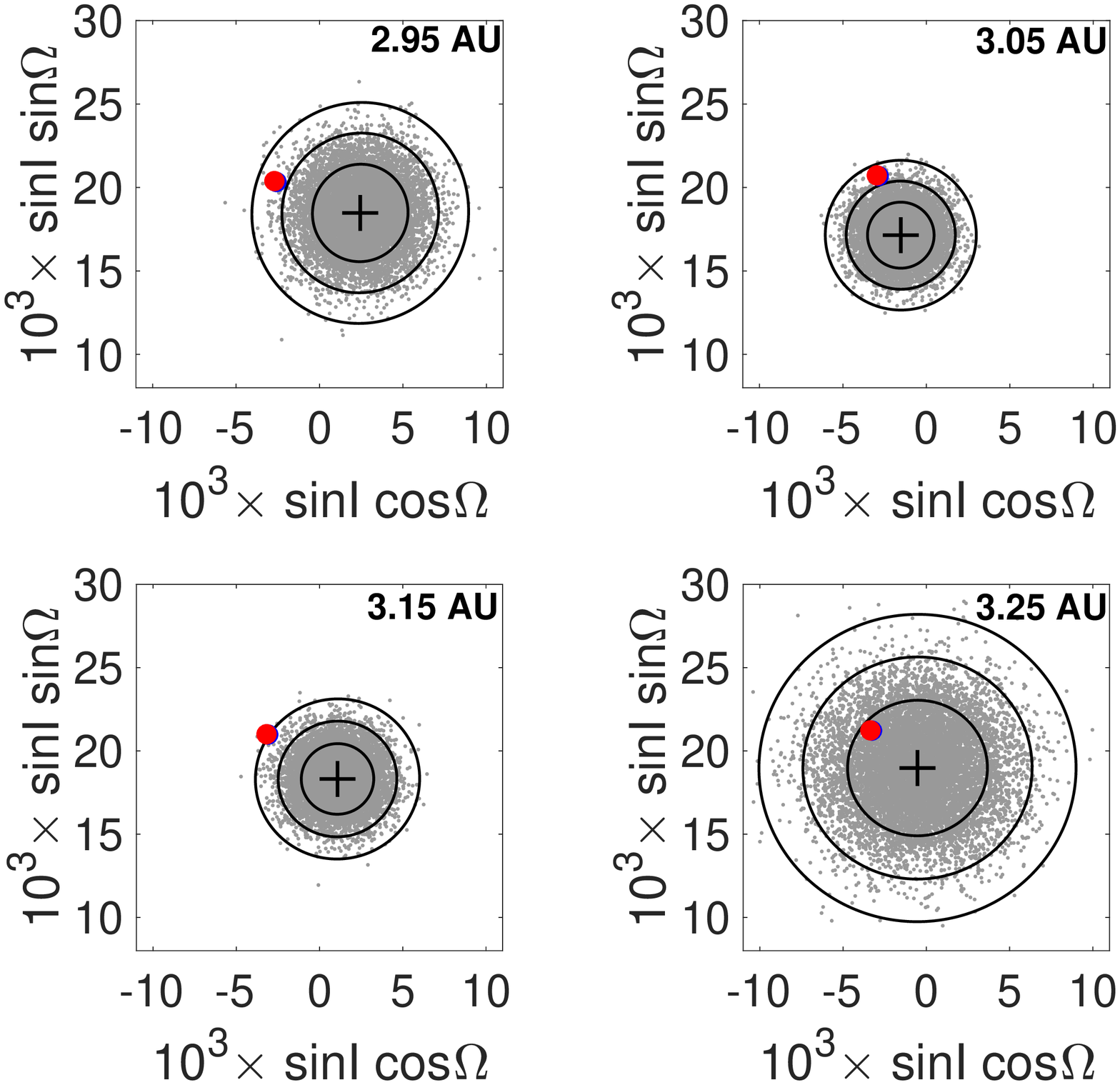}
\end{center}
\caption{Similar to Figure~\ref{inner}, but for the outer belt MBAs.}
\label{outer}
\end{figure*}

In each bin, the methods presented in Sections \ref{s:method1} and \ref{s:method2} are applied to measure the mean plane and its measurement uncertainty.
The comparison between the measured mean plane and the prediction of the secular theory is reported in Figure \ref{a_warps}, which plots the ecliptic inclination and longitude of ascending node in the top panel and the bottom panel, respectively. The numerous asteroids in the bins above 2.10 AU allow a good measurement of the mean plane. By means of a first visual analysis, the minimum in the predicted forced inclination at roughly 2.20 AU is resolved by the measured values which lie close to the theoretical curves and have relatively small measurement uncertainty. The measured ecliptic inclination of the mean plane for 2.40--2.50 AU, however, seems to deviate from the predicted value by more than 1--$\sigma$; this deviation is the starting point of what seems a systematic behavior of the measured mean plane's ecliptic inclination: each semi-major axis bin in the far inner belt, middle belt and outer belt has a mean plane inclination visibly slightly lower than that predicted by the secular solution. The longitude of ascending node of the measured mean planes seems to follow the predicted pattern from the far-inner belt up to the beginning of the outer belt; exceptions to this behavior are recorded at the inner edge, 2.10 and 2.20 AU, where the measured value departs from the secular solution, and also in the outer belt, between 2.80 and 3.30~AU, where the measured values seem to oscillate about a value which is not the predicted value. 

For each bin beyond 2.10 AU, the statistical significance of the deviations from the theoretically expected mean plane is studied in Figure \ref{inner} (semi-major axis between 2.10 and 2.50 AU), Figure~\ref{middle} (semi-major axis between 2.50 and 2.90 AU) and Figure~\ref{outer} (semi-major axis between 2.90 and 3.30 AU). The local measured mean plane is indicated with a black cross on the ecliptic $(Q,P)$ plane; the forced inclination vector computed from the linear secular solution is indicated by a blue dot, while the red dot indicates the forced inclination vector computed with the numerical solution for the planets \citep{Morbidelli:2002}). The 1-$\sigma$, 2-$\sigma$, 3-$\sigma$ contours of the bootstrapped mean planes are also reported. In each panel in these figures, the abscissa and ordinate have aspect ratio of unity, so the uncertainty contours have their true shape.  We note that these contours are nearly circular; this is because the mean poles are not very far from the ecliptic pole. The relative sizes of the uncertainty contours in the different panels are related to the dispersion of the orbit planes about their mean plane and to the sample size; the former does not vary greatly across the semi-major axis bins, but the latter does, so the sizes of the uncertainty contours are driven mainly by the sample sizes.
Examining these results, we observe that for most of the inner and middle belt, specifically the semi-major axis range 2.20--2.90~AU, the theoretically expected mean poles are located within 2--$\sigma$ of the measured mean pole. In the outer belt (semi-major axis between 2.90 and 3.30 AU) the deviations are larger but all within the 3--$\sigma$ contour; the bin centered at 3.15~AU shows the largest deviation, for which the predicted mean poles are located on the 3--$\sigma$ contour. The $(Q,P)$ projections of the mean poles show that the systematic trend with semi-major axis that is visible in Figure \ref{a_warps} has low statistically significance, as the measurement uncertainties for most semi-major axis bins are larger than the measured poles' deviations from the predictions of secular theory.  The most statistically significant deviation is found in the bin centered at 2.15~AU, for which the predicted mean poles are located well outside the 3--$\sigma$ uncertainty contour of the measured mean pole.

\begin{figure}
\includegraphics[width=1\linewidth]{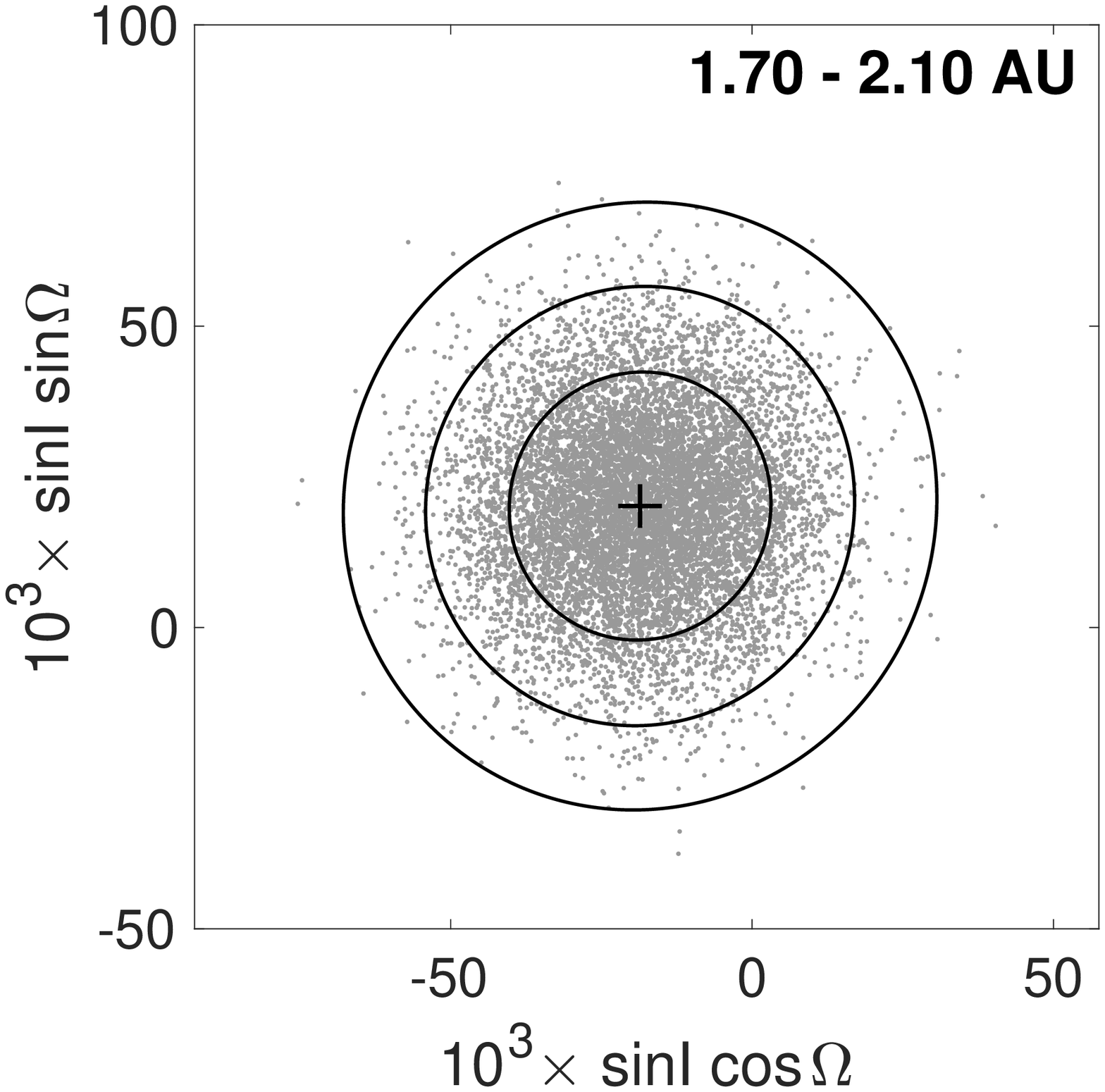}
\caption{The mean pole (black cross) of the pared sample of MBAs with semi-major axis smaller than 2.10 AU is indicated by the black cross; the gray dots are the bootstrapped population of mean poles projected onto the $(Q,P)$ plane. The concentric black contours are the ellipses containing the 68.3\%, 95.4\% and 99.7\% of the population, respectively.}
\label{below210AU}
\end{figure}

At the inner edge of the asteroid belt, our sample in the semi-major axis range 1.70--2.10 AU has only 431 MBAs, much smaller than the sample sizes in the 0.1~AU wide bins we considered above for the more distant MBAs.  We therefore consider this sample as a single bin.  We find that this sample has a mean pole of inclination $1.6 \pm 1.3$ degrees and longitude of ascending node of $133 \pm 47$ degrees. We plot in Figure \ref{below210AU} the projection of the mean pole onto the ecliptic $(Q,~P)$ plane (shown by the black cross) as well as the scatter of the bootstrapped mean poles. The forced inclination predicted by the secular solution has a large variation across this semi-major axis range, but the available sample size here is too small and too dispersed in inclinations to meaningfully compare the theory with the observations.

\section{Summary and Discussion}
\label{s:discussion}

For the main belt asteroids, taken as a whole, we find that their mean plane has ecliptic inclination $\bar{I}=0.93\pm0.04$ degrees and longitude of ascending node $\bar{\Omega}=87.6\pm2.6$ degrees.  Our measurement deviates significantly (more than 5--$\sigma$) from previously published estimates of \cite{Plummer:1916} and \cite{Shor:1992}, both of which undoubtedly suffered from a small sample size compared to what is available today, and effects of collisional families. The measured mean plane is also clearly distinguishable from the solar system's invariable plane as well as Jupiter's orbit plane. 

Because the major planets' orbits are not exactly co-planar, linear secular perturbation theory predicts that the mid-plane of the asteroid belt is not quite a flat sheet, but has a significant warp near its inner edge due to the $\nu_{16}$ nodal secular resonance.  To test this prediction, we measured the mean plane of asteroids binned in 0.1~AU--wide semi-major axis bins. The sample sizes in most bins are large enough that we can confirm the variations of the mid-plane with semi-major axis. We find that the measured mean plane in most bins is consistent (within 3--$\sigma$ significance level) with the prediction of secular theory. A weak systematic trend of the deviations with semi-major axis that appears to be present (Figure~\ref{a_warps}) is not statistically significant with the present sample size. The magnitude of the deviations indicates that a decrease of the measurement uncertainties by a factor of $\gtrsim3$ would allow this trend to be tested more robustly.  This means that a sample size increase by a factor of $\gtrsim9$ is needed for such a test. The forthcoming survey by the LSST is expected to increase the MBAs sample by a factor of $\gtrsim 10$ \citep{Ivezic:2014} and would possibly allow such a test to be feasible.

One prominent exception is the bin near the inner edge of the asteroid belt, centered at 2.15 AU, for which the predicted mean pole is located well outside the 3--$\sigma$ uncertainty range of the measured mean pole. This bin is close to the location of the $\nu_{16}$ nodal secular resonance, hence has significant variation of the forced inclination vector predicted by the secular solution. It is also in close proximity to the $\nu_6$ apsidal secular resonance and the 4:1 mean motion resonance with Jupiter, both of which excite orbital eccentricities of asteroids to high values and are the reason for the relatively low population of asteroids in this region \citep{Michel:2000}. Despite the smaller sample size, it is clear that the mean plane of asteroids in the semi-major axis bin 2.10--2.20 AU has a statistically significant deviation from the forced inclination vector predicted by the linear secular solution as well as the numerical secular solution.  We conclude that secular theory is not sufficient to identify accurately the mid-plane of asteroids in this region. 

We also find a noteworthy deviation of the measured mid-plane from the predicted mid-plane in the bin centered at 3.15 AU; this deviation is just about 3--$\sigma$ significance level. This bin is in proximity to the 2:1 mean motion resonance with Jupiter.

We suspect that the explanation for the above two discrepancies between theory and observation is the breakdown of the secular theory in the vicinity of the 4:1 and the 2:1 Jovian mean motion resonance; the effects of mean motion resonances on the local forced inclination is not accounted for in the secular theory. One might ask: why do we not see similarly significant discrepancies in the vicinity of other Jovian mean motion resonances located in the main asteroid belt, such as the 3:1, 5:2 and 7:3? The answer is that these are the locations of the prominent Kirkwood gaps where very few asteroids are present and therefore do not significantly contribute to the local measured mid-plane.  Investigating the effects of mean motion resonances on the forced inclination is warranted for a better understanding of the dynamical structures in the asteroid belt; such an investigation is beyond the scope of the present work. 

Our measurement of the mid-plane of asteroids between 1.70 AU and 2.10 AU has a significantly higher measurement uncertainty than that of the other semi-major axis bins. This is because of the small number of asteroids in this region, and also because the population has a peculiar inclination distribution, with a prominent concentration near 20 degrees. Theoretical studies of this region of the asteroid belt have identified a complex dynamical structure, threaded with overlapping mean motion resonances with Jupiter and Mars as well as non-linear secular resonances \citep{Milani:2010, Michel:1997, Michel:2000}. We note that the Hungaria family of asteroids -- which exists in this region and which we removed for our pared sample -- has an uncertain genesis~\citep{Milani:2010}.  If we include the Hungarias (of absolute magnitude $H<15.5$), the sample size increases to about 700, but it does not significantly change the measured mid-plane nor its measurement uncertainty. Future investigation of the mid-plane in this interesting region would benefit from the availability of a larger observationally-complete sample.

We comment on the magnitude and sources of the measurement uncertainty of the mean pole of MBAs. The 1--,2-- and 3--$\sigma$ uncertainties of the overall mean pole (angular distance between the mean pole and the 1--,2-- and 3--$\sigma$ circle) of the asteroid belt are 0.04, 0.07 and 0.1 degrees. The 1--$\sigma$ uncertainties of the mean pole locations of the binned samples in the semi-major axis range 2.10--3.30 AU are 0.1--0.2 degrees.
A first source of uncertainty of the measured mean pole location is the random error, which is due to the finite size of the sample. As the sample size increases, the standard error tends to decrease as $\sim N^{-\frac{1}{2}}$ according to the central limit theorem. One way to increase our sample of MBAs is to include the unnumbered minor planets in the set. Most of these minor planets, however, are faint objects (near the current limits of observational completeness) and some of them may also be unrecognized collisional family members; their inclusion in the set must be done carefully, since it is likely to reinforce the level of systematic error, which is the second source of error. 
A source of systematic error is the presence of unrecognized family members in our sample of MBAs. 
The AstDys-2 catalog of families has been generated using conservative criteria for family membership, in order to minimize the number of false groupings and to reduce the chance of spurious merged families that can occur given the very large sample size; this can potentially lead to an underestimate of family relationships \citep{Migliorini:1995,Milani:2014}. We note, however, that differential nodal precession of family members will tend to randomize the nodes of their orbital planes around the local forced plane, so only the relatively recently-formed family clusters--those whose nodes have not fully randomized--will contribute significantly to the systematic error of the local mean pole measurement.

Recently, \cite{Delbo:2017} reported the discovery of an ancient collisional family in the inner asteroid belt which overlaps with the previously identified Polana family; it includes about 100 asteroids not previously linked to any other family.  We re-computed the mean poles by removing from our pared sample the newly identified collisional family members; we found that the positions of the mean poles are not significantly affected (the differences are much smaller than 1--$\sigma$).

An additional potential source of systematic bias is the so-called "ecliptic" bias, that is, most observational surveys are carried out close to the ecliptic, so the measurement of the mean pole as the average of the orbit pole unit vectors causes a measurement bias toward smaller ecliptic inclination of the mean pole \citep{Knezevic:1982}. To check this, we also computed the mid-plane of MBAs as defined by the plane of symmetry of the transverse component of their heliocentric velocity vectors (we refer to \citep{Volk:2017} for more details about this method). This method is much less sensitive to non-uniform sky locations of the observational sample of MBAs. The mean poles measured with this method (for all semi-major axis bins) are within the 2--$\sigma$ uncertainty ellipses of those calculated with the method of averaging the orbital pole unit vectors, and there is no systematic trend of the average-of-pole-vectors being closer to the ecliptic than the plane-of-symmetry.  We conclude that the "ecliptic bias" is not significant for this sample.

We also recognize that our pared sample contains some asteroids that are strongly chaotic and unstable on long timescales. For example, the population of Mars-crossing asteroids are understood to have their orbital elements change chaotically due to close encounters with Mars \citep{Michel:2000}.  Such asteroids in our sample may contribute to the random and systematic error of the measured mean pole.  We re-computed the mean poles by removing from our pared sample all the Mars-crossing asteroids (that is, those whose perihelion distance is less than Mars' aphelion distance); we found no statistically significant differences in the results.

As large-sky observational surveys, such as the ongoing Pan-STARRS\footnote{https://panstarrs.stsci.edu/} and the Catalina Sky Survey\footnote{https://catalina.lpl.arizona.edu/} and the forthcoming LSST survey\footnote{https://www.lsst.org/}, increase the census of asteroids to fainter magnitudes, the observationally complete sample size will increase and thereby allow more accurate measures of the mean plane of the asteroid belt.  It would be particularly interesting to revisit with a larger sample size the mean plane measurements near the $\nu_{16}$ warp in the inner asteroid belt as well as those near the 2:1 Jovian mean motion resonance in order to shed light on the origin of the deviations from secular theory; a larger sample near the inner edge of the asteroid belt would also help to better resolve the rapid variation of the expected mean plane in this region and would allow a better test of the secular theory with observational data. \\

The data used in this paper and our computer codes are available upon request.

\section*{Acknowledgement} 
We are grateful for funding from NSF (grant AST-1312498). We thank Kat Volk and Matthew Hedman for discussions, and the reviewer for comments that improved this paper.

\end{document}